\documentclass[twocolumn,aps]{revtex4-1}
\usepackage[latin9]{inputenc}
\usepackage{textcomp}
\usepackage{amsmath}
\usepackage{amssymb}
\usepackage{graphicx}

\makeatletter

\DeclareFontEncoding{LGR}{}{}
\DeclareRobustCommand{\greektext}{%
  \fontencoding{LGR}\selectfont\def\encodingdefault{LGR}}
\DeclareRobustCommand{\textgreek}[1]{\leavevmode{\greektext #1}}
\ProvideTextCommand{\~}{LGR}[1]{\char126#1}

\providecommand{\tabularnewline}{\\}

\@ifundefined{textcolor}{}{%
\definecolor{BLACK}{gray}{0}
\definecolor{WHITE}{gray}{1}
\definecolor{RED}{rgb}{1,0,0}
\definecolor{GREEN}{rgb}{0,1,0}
\definecolor{BLUE}{rgb}{0,0,1}
\definecolor{CYAN}{cmyk}{1,0,0,0}
\definecolor{MAGENTA}{cmyk}{0,1,0,0}
\definecolor{YELLOW}{cmyk}{0,0,1,0}
}

\usepackage{upgreek}

\providecommand{\tabularnewline}{\\}

\@ifundefined{textcolor}{}{%
\definecolor{BLACK}{gray}{0}
\definecolor{WHITE}{gray}{1}
\definecolor{RED}{rgb}{1,0,0}
\definecolor{GREEN}{rgb}{0,1,0}
\definecolor{BLUE}{rgb}{0,0,1}
\definecolor{CYAN}{cmyk}{1,0,0,0}
\definecolor{MAGENTA}{cmyk}{0,1,0,0}
\definecolor{YELLOW}{cmyk}{0,0,1,0}
}

\usepackage{color}%\bibliographystyle{plain}
\def\NOT(#1,#2){\OneQubitGate(#1,#2){$X$}}

\@ifundefined{showcaptionsetup}{}{%
\PassOptionsToPackage{caption=false}{subfig}}

\makeatother

\begin{document}

\title{CVD-growth of ultra-pure diamond, generation of NV centers by ion-implantation
and their spectroscopic characterization for quantum technological
applications}

\author{T. Chakraborty$^{1,2}$, F. Lehmann$^{1}$, J. Zhang$^{1}$, S. Borgsdorf
$^{3}$, N. Wöhrl$^{4}$, R. Remfort$^{4}$, V. Buck$^{4}$, U. Köhler$^{3}$,
D. Suter$^{1}$ }

\affiliation{$^{1}$Fakultät Physik, Technische Universität Dortmund, D-44221
Dortmund, Germany}

\affiliation{$^{2}$Institute for Materials Research (IMO), Hasselt University,
Wetenschapspark 1, B-3590 Diepenbeek, Belgium}

\affiliation{$^{3}$Experimentalphysik IV, AG Oberflächen, Ruhr-Universität Bochum,
Germany}

\affiliation{$^{4}$Faculty of Physics, University Duisburg-Essen and CENIDE,
Germany}
\begin{abstract}
Applications of nitrogen-vacancy (NV) centers in diamond in quantum
technology have attracted considerable attention in recent years.
Deterministic generation of ensembles of NV centers can advance the
research on quantum sensing, many-body quantum systems, multipartite
entanglement and so on. Here we report the complete process of controlled
generation of NV centers in diamond as well as their characterisation:
growing diamond films through chemical vapor deposition (CVD), ion
implantation and spectroscopic characterization of the defect centers
using a confocal microscope. A microwave-assisted CVD set-up is presented
which we constructed for the preparation of single-crystalline homoepitaxial
diamond films. The films were prepared with minimized nitrogen concentration,
which is confirmed through photoluminescence measurements. We demonstrate
an in situ ultra high vacuum (UHV) implantation and heating process
for creation of NV centers using a novel experimental set-up. For
the first time hot implantation has been shown which prevents surface
charging effects. We do not observe graphitization due to UHV heating.
By optimizing the implantation parameters it has been possible to
implant NV centers in a precise way. We present large area mapping
of the samples to determine the distribution of the centers and describe
the characterization of the centers by spectroscopic techniques. Reducing
the decoherence caused by environmental noise is of primary importance
for many applications in quantum technology. We demonstrate improvement
on coherence time $T_{2}$ of the NV spins by suppression of their
interaction with the surrounding spin-bath using robust dynamical
decoupling sequences. 
\end{abstract}
\maketitle

\section{Introduction}

Quantum technology exploits a handful of solid state systems whose
physical properties are determined by quantum effects. The nitrogen-vacancy
(NV) color center, a defect center consisting of a substituted nitrogen
atom and an adjacent vacancy embedded in a diamond crystal \citep{Jelezko_review},
is one of them. It has several attractive properties, including millisecond-scale
spin coherence time, the possibility of manipulating the spins through
microwave (MW) pulses, efficient optical initialization and detection
of the spins and the ability to perform such experiments at room temperature
\citep{Jelezko_PRL,Dutta_Science,chakraborty2017polarizing}. These
properties make it a useful solid state qubit system which has already
been applied in a number of important quantum information experiments
like demonstrating long quantum memory by controlling the spin-qubits
with high fidelity \citep{Maurer1283}, coherently manipulating individual
nuclear spins by specifically addressing a proximal electronic spin
\citep{Childress_Science}, exhibiting quantum entanglement between
a photon and solid-state spin \citep{Togan_Nature_466_730_2010},
demonstrating quantum interference between two photons \citep{Sipahigil_2012,Bernien_PRL},
implementation of a quantum memory \citep{Fuchs_2011}, and quantum
repeater \citep{Childress_PhysRevLett.96.070504}. On the other hand,
there has been a significant amount of progress in NV-based sensing
technologies where NV centers are applied in imaging the dynamics
of neural network \citep{NJP-13-4-045021}, living cells \citep{liu2016fluorescent,Balasubramanian_455_10},
probing of superconducting effects \citep{1367-2630-13-2-025017},
sensing magnetic fields with high precision at the nanoscale \citep{maze2008nanoscale,taylor2008high,grinolds2013nanoscale}
and others. Such promising applications have pioneered a way towards
state-of-art NV-based quantum technology.

These sensing applications can be compared to other technological
advancements like superconducting quantum interference devices \citep{McDermott25052004},
magnetic resonance force microscopy \citep{rugar2004single}, Hall
probe microscopy \citep{Chang:1992aa}, optical atomic magnetometer
\citep{Xu22082006} have been introduced in the past decades concerning
sensing and imaging of magnetic field. Comparison shows that the novel
method of magnetic field detection using the NV centers has several
advantages over the more conventional techniques. For instance, the
NV centers in diamond crystal can be prepared in a controlled fashion:
the implantation process can be adjusted to reach a required density
of NV centers at a depth of a few nanometers from the surface. This
is an essential condition for efficient sensing as such shallow centers
can be brought in close proximity ($\sim$nm) to the magnetic centers
under investigation. Importantly, diamond-based sensing can operate
at room temperature, in liquid media and normal atmospheric conditions
which gives more flexibility and feasibility to such applications.
It has been experimentally demonstrated that diamond spins can detect
weak magnetic field with high sensitivity and nanometer scale spatial
resolution \citep{maze2008nanoscale,taylor2008high,Balasubramanian_455_10}. 

The sensitivity of NV-based sensors is generally higher if ensembles
of centers are used, rather than single centers. Sensing with dense
ensembles of NV centers can increase the signal-to-noise ratio as
the generated photoluminescence signal increases with $N$, the number
of centers being used for detection. Thus, the magnetic sensitivity
can be enhanced by a factor $1/\sqrt{N}$ when the field is homogeneous
over the magnetometer area \citep{taylor2008high}. Furthermore, ensembles
of centers are capable of imaging magnetic field over a wide field-of-view
range \citep{hong2013nanoscale}. 

Ensemble of centers has already been applied in different contexts
like quantum metrology \citep{Acousta_PhysRevB.80.115202,NJP-13-4-045021},
quantum walk \citep{Hardal_PhysRevA.88.022303}, device physics \citep{zhu2011coherent,zhu2014observation,putz2014protecting},
and quantum simulations \citep{PhysRevA.86.012307} . In this context,
there have been a significant amount of efforts regarding growing
highly-pure diamond films and implanting clusters of NV centers with
a target pattern suitable for the desired applications.

The primary motivation of this project is to generate ensembles of
NV centers in ultra-pure single crystal diamond films in a controlled
way with the goal of optimising them for applications in quantum information
and sensing technology. The content of the paper has been mainly divided
into three parts: the first part deals with growing highly pure single
crystalline diamond films. The second part describes the precise creation
of near-surface NV centers by ion implantation. In the third part
we describe experimental methods for characterizing the centers and
we also demonstrate experiments using various dynamical decoupling
sequences with an aim to reduce the dephasing of the NV spins due
to the environmental noise.

\section{Homoepitaxial Growth of Diamond}

\begin{figure}
\includegraphics[width=1\columnwidth]{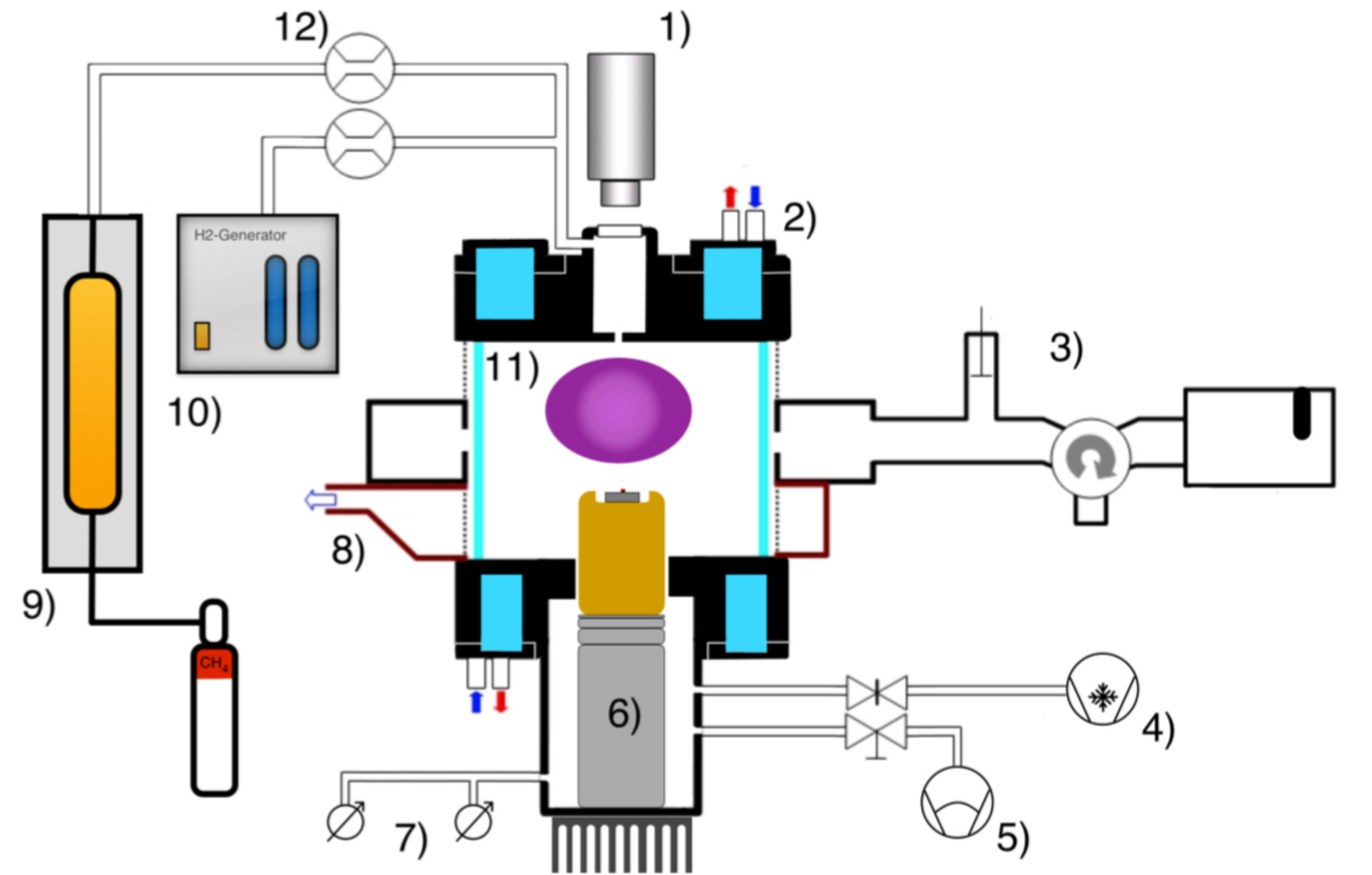}\caption{Schematic representation of the experimental setup: 1) infrared pyrometer
IS320 in line of sight with sample 2) water cooling in top and bottom
flange 3) microwave feed consisting of magnetron, circulator, EH tuner
and ring resonator 4) cryopump with plate valve 5) backing pump 6)
multi-part sample holder 7) vacuum gauges 8) air cooling 9) methane
gas purifier MonoTorr PS4-MT3-531 10) hydrogen generator HG1200 11)
quartz cylinder 12) mass flow controller. \label{fig:PCVD}}
\end{figure}

The microwave-activated plasma process \citep{Kamo1982} has been
established as a more efficient technique for the deposition of high-quality
single-crystal diamond films than other existing diamond growing techniques
like HPHT (high pressure high temperature) crystal growth \citep{bundy1955man,Liander1955}
or the hot-filament method \citep{Matsumoto1981}. In this work, pure
single-crystal diamond films are homoepitaxially grown by chemical
vapor deposition in a microwave assisted CVD plasma source (CYRANNUS®
by iplas). We have developed and optimized an experimental set-up
for growing diamond films with minimum nitrogen concentration. A schematic
diagram of our set up is shown in Fig. \ref{fig:PCVD}.The technology
of the plasma source CYRANNUS® is based on a cylindrical resonator
with annular slots. The microwaves are fed into the ring resonator
by a waveguide from which it is coupled into the cylindrical resonator
of the reactor. Due to the centric position of the plasma in the reactor,
the excitation without electrodes and the short mean free path at
the deposition pressure of ca. 200 mbar the interaction of the reactive
gas with chamber walls is minimized. 

The homoepitaxial growth of diamond by CVD involves the deposition
of carbon atoms on a diamond surface that originate from the dissociation
of carbon-containing precursors in hydrogen gas. Hydrogen is essential
because it selectively etches non-diamond carbon. For the deposition
of pure diamond films the hydrogen in this work is generated in a
hydrogen generator (HG-1200 by CMC Instruments GmbH) with a purity
of 99,99999\%. The methane (5.0) is purified (saes Pure Gas Inc.)
to gas impurities \textless 1 ppb. Diamond films are grown with a
methane and hydrogen mixture with a typical ratio around 4\%).

The chamber is evacuated with a scroll pump and a cryopump (CTI-Cryogenics
8200 compressor with cryo-torr pump) and reaches a base pressure of
around 10$^{-9}$ mbar. Mirror-polished single crystal Ib (001) diamonds
(electronic grade) from Element Six Ltd. (\textless{} 5 ppb nitrogen
impurities) were used as substrates. Substrates were ultrasonically
cleaned for 20 min in isopropanol and subsequently 20 min in acetone.
During deposition the substrates were kept on a copper substrate holder.
Prior to the deposition the substrates were etched in a H$_{2}$ plasma
for 30 min at 170 mbar. After this cleaning step the process parameters
were switched to the deposition parameters (Table \ref{tab:Relevent-process-parameters}).

The thickness of the grown diamond films and comparison between purity
of the substrate and films were determined by measuring PL depth profiles,
as described in details in section IV.B. 

\begin{table}
\begin{tabular}{|c|c|}
\hline 
Parameters  & Values\tabularnewline
\hline 
\hline 
Pressure  & 170 mbar\tabularnewline
\hline 
Gas flow  & 400 sccm\tabularnewline
\hline 
H$_{2}$ fraction  & 96\%\tabularnewline
\hline 
CH$_{4}$fraction  & 4\%\tabularnewline
\hline 
MW-power  & 1.6 kW\tabularnewline
\hline 
Deposition time  & 8 hrs.\tabularnewline
\hline 
Substrate  & Diamond substrates Ib (001) (electronic grade)\tabularnewline
\hline 
Temperature  & 900$^{0}$ C\tabularnewline
\hline 
\end{tabular}\caption{Relevant process parameters for the grown samples \label{tab:Relevent-process-parameters}}
\end{table}

\begin{figure}
\includegraphics[width=1\columnwidth]{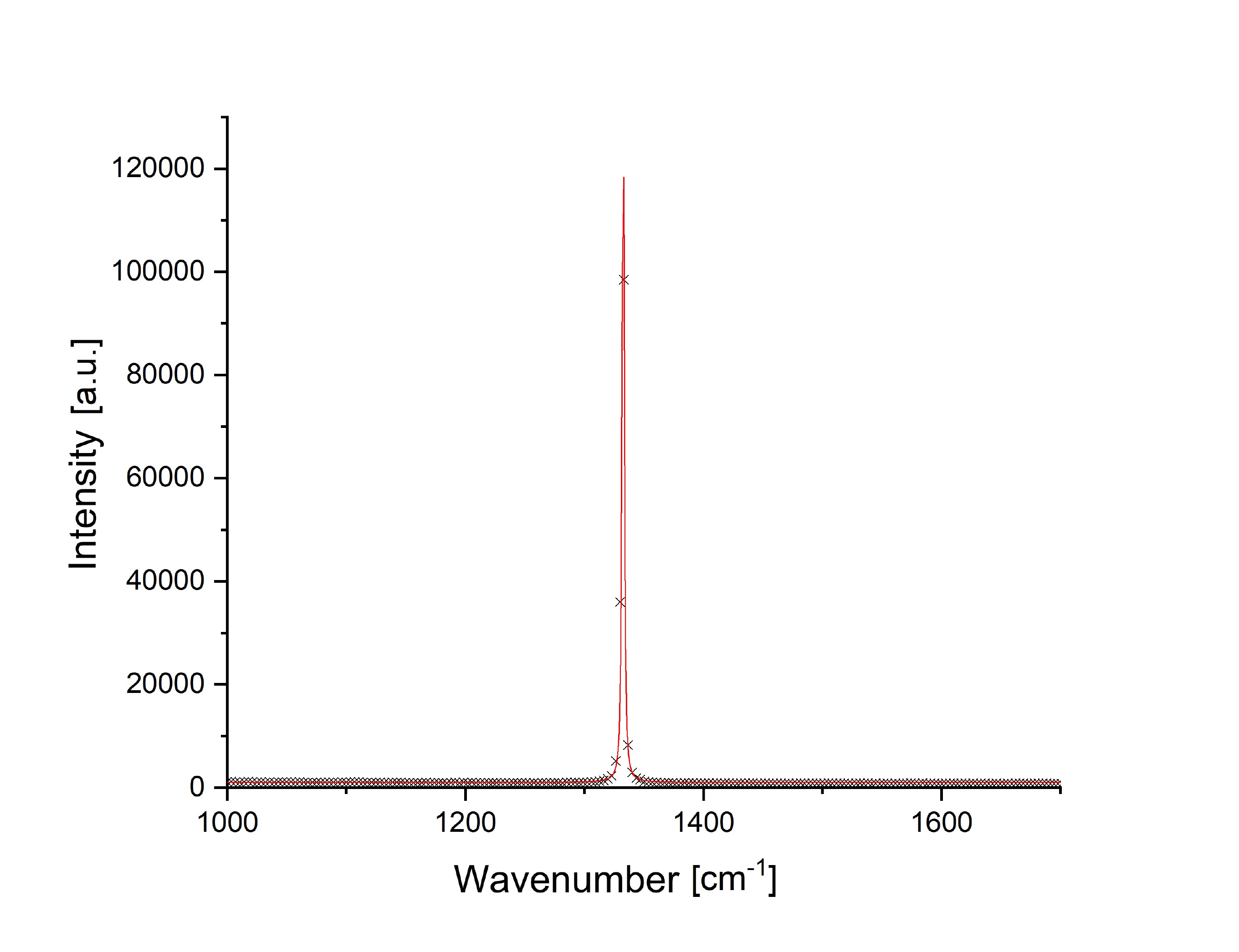}\caption{Raman spectroscopy of single crystal diamond. }
\label{EG3-06_raman}
\end{figure}

Raman spectroscopy was performed to determine the crystallinity of
the synthesized diamond samples. The data (Figure \ref{EG3-06_raman})
show a sharp diamond peak at 1332.54 cm$^{-1}$ with a FWHM of 1.61
cm$^{-1}$ . The narrow line proves the high crystallinity and even
exceeds the values for natural and HPHT (high pressure high temperature)
single crystals \citep{shu2017epitaxial}. The position of the diamond
peak indicates that there is almost no intrinsic stress in the grown
films. Similar spectra were obtained over the whole surface area.

Secondary Ion Mass Spectroscopy (SIMS) measurements were done to measure
the impurities in the deposited diamond films. Since impurities were
expected to come from the process gas, we measured the concentration
of hydrogen, oxygen and nitrogen in the films. For hydrogen and oxygen
the signals were below the detection limits of the instrument. The
upper limit of nitrogen concentration was determined using the approach
of Gnaser \textit{et al.} \citep{gnaser2001greatly} as 1.3$\cdot$10$^{-7}$(ca.
0.1 ppm). The only other element that was found in the diamond sample
in relevant concentrations is silicon. It was measured in PL measurements
with a strong ZPL at 738 nm originating from the silicon-vacancy center
in diamond. The silicon is supposed to come from the plasma reactor
walls since the vacuum chamber consists of a quartz-recipient and
the plasma has contact with the wall and etches silicon from the quartz.
It was not possible to calculate the concentration of silicon from
the PL measurements. Future SIMS measurements will include the determination
of the concentration. 

The incorporated nitrogen can only come from the process gases or
the leakage of the vacuum system. By using a hydrogen generator (cmc
Instruments GmbH type HG1200-2T) and pure methane which was additionally
purified (Mono Torr type PS4-MT3-531), the purity of the process gases
used was 7.0 (99.99999\%) for hydrogen and 9.0 (99.9999999\%) for
CH4, while the leakage of the vacuum chamber was estimated to be around
2.4\ensuremath{\centerdot}10$^{-4}$ sccm. The nitrogen concentration
of the grown layer is thus calculated to be below 1 ppb by assuming
an incorporation rate of 10$^{-4}$\citep{jin1994effect}.

\section{Ion Implantation}

Two processes are currently used to generate NV centers in diamond.
One process is to dose nitrogen in the process gas while growing the
CVD diamond. In this process the NV centers are statistically distributed
in a well defined plane. It is possible to tune the thickness of this
plane \citep{balasubramanian2009ultralong} and the density of NV
centers. The second process is to implant nitrogen with an ion gun
into the single crystal diamond \citep{meijer2008towards}. For scalable
quantum computer architecture and also for magnetometry applications,
it is necessary to get control of the accurate positioning of the
NV centers in all three axes together with a high probability of creation.
It was shown that the yield of creating NV centers can be up to 50\%
with high implantation energies (2 MeV) \citep{meijer2005generation}
but the spatial resolution is very low. The depth and straggling depends
on the ion energy and can be calculated with SRIM \citep{acosta2009diamonds}.
It is possible to increase the resolution by implanting with lower
energies (1-5 keV) but in this case the yield of creation is reduced
to 2.5\% \citep{rabeau2006implantation}. Also, the ion density has
an influence on the yield \citep{pezzagna2010creation}, therefore
it is important to get control of the beam current, the beam profile
and the implantation time. To create the NV centers it is necessary
to anneal the diamond after the implantation up to 600$^{\circ}$
- 800$^{\circ}$ C for 2 hours which is typically done under high
vacuum conditions in a quartz oven after the diamond has been transferred
from the implantation chamber. The annealing of the diamond under
these conditions often leads to surface graphitization, which must
be removed by etching in a boiling triacid \citep{cui2013effect}. 

The diamond surface before and after the implantation is still not
well characterized, in particular the influence of the surface on
shallow NV centers. To check how treatments influence the quality
of the implantation, e.g. lead to a higher background in the confocal
signal or harm the coherence time $T_{2}$ time we want to present
a completely new all in ultra high vacuum (UHV) setup where the preparation
of the diamond surface, the implantation and annealing process can
be done in situ under UHV conditions. Additionally, only in an UHV
environment completely non terminated diamond surfaces can be prepared.
As received diamonds from Element6 are typically terminated by oxygen
because the samplewas cleaned in boiling triacid after the CVD growth.
If the diamond is not cleaned in this way, it is hydrogen-terminated
after the CVD growth. The influence of oxygen and hydrogen on the
ratio NV$^{-}$/NV$^{0}$ has been investigated in detail \citep{grotz2012charge}
and it was also found that a hydrogen terminations leads to a surface
conductivity on diamond. Our setup gives us for the first time the
chance to check systematically the influence of these typical surface
terminations on the creation yield for NV centers by comparing them
to a non-terminated surface. We use a conventional UHV chamber with
a turbo molecular pump, an ion getter pump and a titanium sublimation
pump to achieve a base pressure of 5$\cdot$10$^{-11}$ mbar. 
\begin{figure}[h]
\centering \includegraphics[width=1\columnwidth]{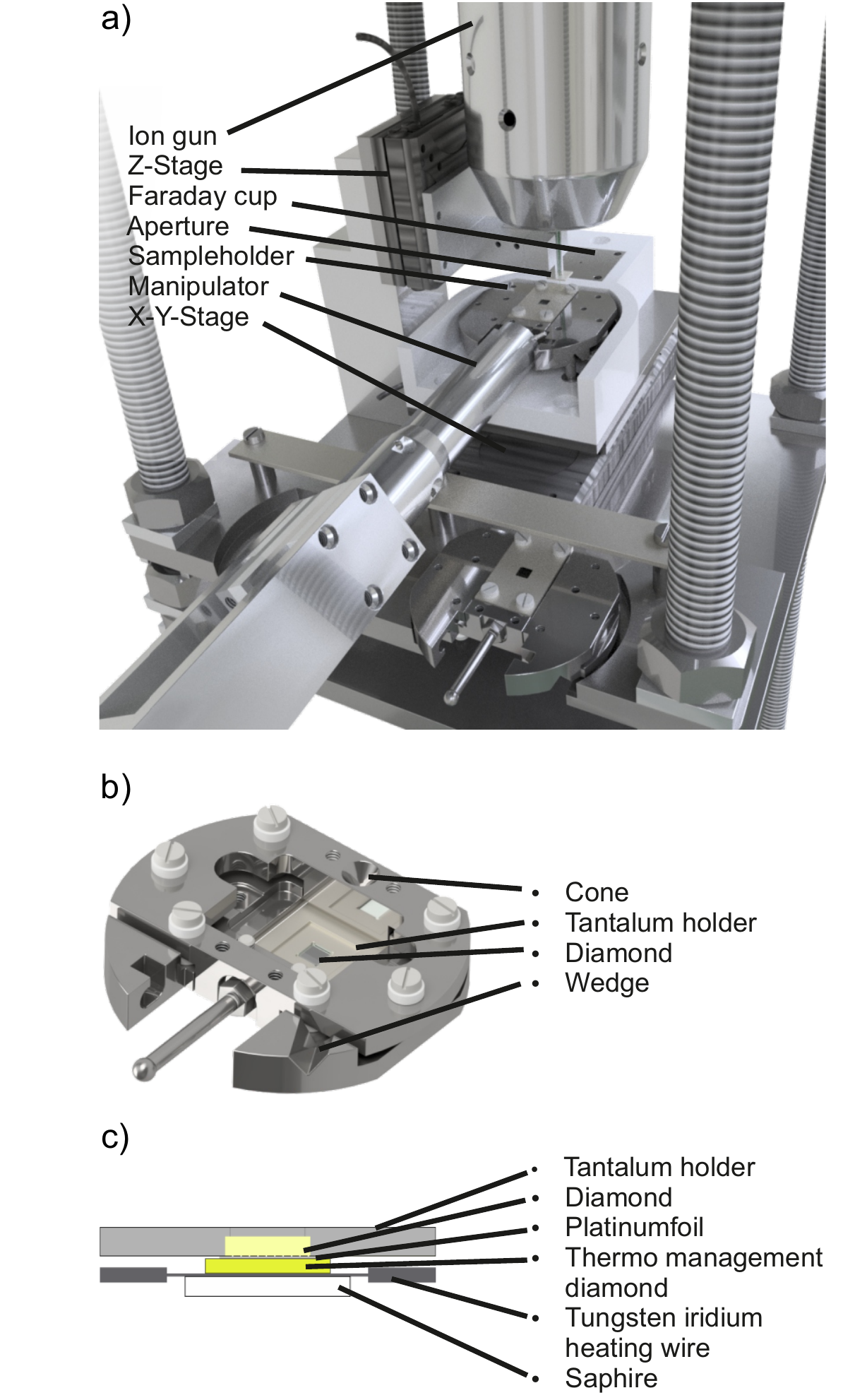} \caption{a) Complete implantation stage. b) Sampleholder for precise placing.
c) Heating stack for effective heat transport.}
\label{fig:aufbau1} 
\end{figure}

For low-energy nitrogen implantation we use a \textit{IQE 12/38} ion
source from \textit{SPECS}. The energy range of the ion gun goes from
400 eV up to 5 keV. The beam diameter (FWHM) is specified to be 150
$\mu$m. To get a very high purity of nitrogen a Wien-Mass filter
is installed, which can also be used to switch between molecular or
atomic nitrogen. As a source gas we use $^{15}$N$_{2}$ with a purity
of (98 atom \%). $^{15}$N nitrogen is used to ensure that the NV
center was created by implanted nitrogen and not from an abundant
one because even with a nitrogen concentration below 5 ppm there would
be a chance of 1 native nitrogen atom in a volume of 150 $nm^{3}$.
To measure the beam current a Faraday cup with shielding is installed.
The beam diameter was calibrated and measured by scanning the beam
over the edge of the Faraday cup. The last parameter which has to
be controlled for a correct dose is the implantation time. An electric
beam chopper allows to switch on and off the ion beam in 1$\mu$s
and to do 15-1500$\mu$s pulses. The sample position is controlled
by an XY-nanopositioning stage from \textit{attoCube} with a closed
feedback loop. To control the width of the implanted region, a tantalum
aperture is used for first test experiments by melting a hole into
the tantalum sheet with a diameter of $\sim$20\,$\mu$m by a laser
beam. The sputter yield for nitrogen on tantalum with 5\,keV is very
low. So it can be assumed that the contamination of diamond by tantalum
is negligible. The aperture is connected to ground to prevent any
charging effect and the focus point of the ion beam is set below the
sample to achieve a more parallel beam path. The position of the aperture
with respect to the sample surface is controlled by a Z-nanopositioning
stage. The sample holder (see Fig.\,\ref{fig:aufbau1} b)) was designed
to transfer the diamond into the UHV chamber and to the implantation
stage below the ion gun (see Fig.\,\ref{fig:aufbau1} a)). The accuracy
of placing the sample is below 5$\mu$m, which was achieved by a cone
and wedge placement design. Also it is important to heat the diamond
in UHV up to 900$^{\circ}$C while maintaining a pressure of 1$\cdot$10$^{-10}$
mbar. To achieve this we have designed an effective heating stack
(see Fig.\,\ref{fig:aufbau1} c)).

\begin{table*}
\begin{tabular}{|c|c|c|c|c|c|}
\hline 
Sample & Precleaning & Termination & Aperture & Dose {[}cm$^{2}${]} & Implantation temperature {[}$^{\circ}$C{]}\tabularnewline
\hline 
\hline 
S1 & isopropanol, acetone  & Hydrogen & no & 1$\cdot$10$^{12}$ & Room temperature\tabularnewline
\hline 
S2 & isopropanol, acetone heated at 800 $^{\circ}$C & bare & no & 1$\cdot$10$^{12}$ & Room temperature\tabularnewline
\hline 
S3 & isopropanol, acetone, heated at 800 $^{\circ}$C & bare & no & 1$\cdot$10$^{12}$ & 700\tabularnewline
\hline 
S4 & isopropanol, acetone, heated at 800 $^{\circ}$C & bare & no & 1$\cdot$10$^{17}$ & 700\tabularnewline
\hline 
S5 & isopropanol, acetone, heated at 800 $^{\circ}$C & bare & yes & 4$\cdot$10$^{15}$ & 700\tabularnewline
\hline 
\end{tabular}\caption{The table shows the main information for the samples S1 - S5. The
post heating for all the samples was done at 800 $^{0}$C for 2 hours.
All experiments were done under UHV conditions.}
\label{Implantation_conditions}
\end{table*}

The thermo-management-diamond has a very good thermal conductivity
of 2000 W/Km at room temperature. Also at 900$^{\circ}$C its conductivity
is significantly higher than that of sapphire. The platinum foil also
collects the thermal radiation of the tungsten iridium heating meander
and does not react with the diamond. With this setup we need 12 W
electric power to reach 900$^{\circ}$C on the diamond. Additionally
a four point Van-der-Pauw measuring station was installed in the UHV-system
to determine the surface conductivity of the diamond after different
surface treatments of a pure heated diamond in UHV. 
\begin{figure}[b]
\centering \includegraphics[width=1\columnwidth]{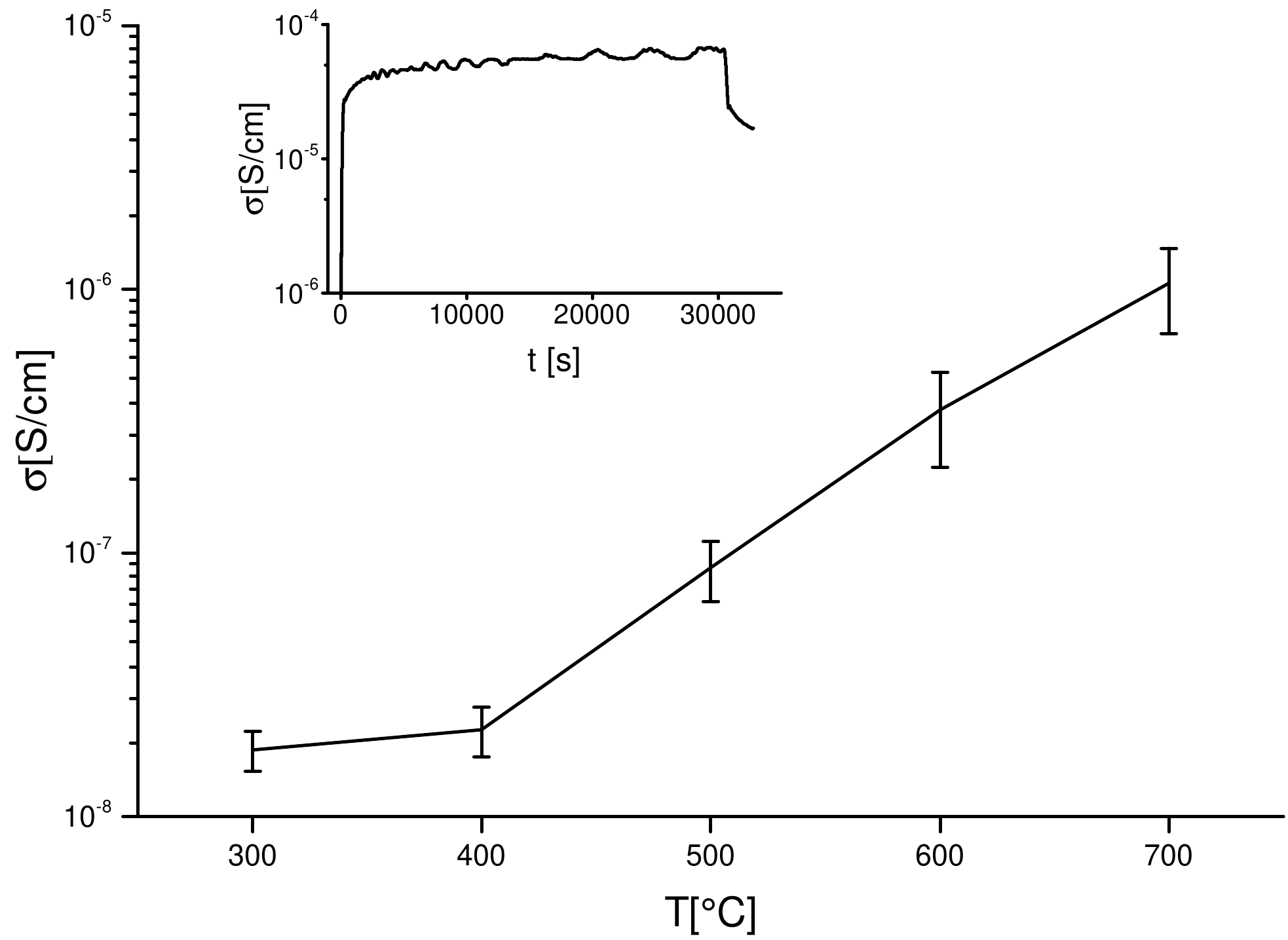} \caption{Van-der-Pauw conductance measurement in UHV. \protect Inset: Time-dependent
measurement of a hydrogen-terminated diamond at room temperature.
The initial fast increase of the conductivity occurs after exposing
the hydrogen-terminated diamond without any adsorbates from vacuum
to atmospheric conditions. After 30000\,s the diamond was placed
again in a vacuum environment, which leads to a drop of the conductivity.
\protect Bottom: Temperature-dependent conductivity of non-terminated
diamond.}
\label{fig:heizmess} 
\end{figure}

Mapping of the diamond samples was performed with a setup without
the XY-Stage and the aperture as it is shown in Fig.\,\ref{fig:aufbau1}.
To check the influence of different surface treatments on the localization
and shape of the implanted area, five implantations were made under
different conditions which are discussed below. Table \ref{Implantation_conditions}
provides an overview of the main sample preparation parameters. Starting
from a base pressure of 5$\cdot$10$^{-11}$ mbar the diffuse N$_{2}$-pressure
during implantation rose to 2$\cdot$10$^{-8}$ mbar. The dose was
1$\cdot$10$^{12}$ cm$^{2}$and the beam was focused at the center
of the sample. They were heated at 800$^{\circ}$C for 2 hours under
a vacuum of 2$\cdot$10$^{-10}$ mbar after the implantation. Sample
1 (S1) was hydrogen terminated and the implantation was done at room
temperature. To check the surface conductivity we use the Van-der-Pauw
setup (see inset of Fig.\,\ref{fig:heizmess}). The hydrogen-terminated
diamond shows a conductance of 1$\cdot$10$^{-3}$ S/m at atmospheric
conditions and it drops to 3$\cdot$10$^{-4}$ S/m in UHV which was
also reported in \citep{kubovic2010electronic}. Sample 2 (S2) was
cleaned before the implantation by heating the sample up to 800$^{\circ}C$
for 12 hours under a vacuum of 2$\cdot$10$^{-10}$ mbar. After this
treatment the sample was no longer hydrogen or oxygen terminated \citep{maier2000origin},\citep{maier2001electron}.Sample
3 (S3) was in the same way cleaned as S2 but the implantation was
done at 700$^{\circ}$C\,$^{0}C$.To check if the heating treatment
leads to a higher surface conductivity we measured the surface conductivity
for different temperatures of a non terminated diamond with the Van-der-Pauw
setup (Fig.\,\ref{fig:heizmess}). It can be seen that the conductance
raises up to 1$\cdot$10$^{-6}$ S/m at 700$^{\circ}$C, which is
more than two orders magnitude lower than for S1 but it seems that
the electrical conductivity is sufficient to prevent charging during
the ion implantation. Optical characterization of the implanted NV
centers in S1, S2 and S3 are discussed in section IV(C). In section
IV(B) we have demonstrated large area confocal mapping of samples
S4 and S5 whose preparation procedures are described below: In the
case of S4, the sample was cleaned with isopropanol and acetone, heated
in UHV (5$\cdot$10$^{-10}$ mbar) at 700$^{0}$C for 2 hours, ion
implantation without an aperture was performed with an ion dose of
1$\cdot$10$^{17}$cm$^{-2}$ and finally heated at 800$^{0}$C for
2 hours. For sample S5, we implanted four dots in a row with a dose
of 1$\cdot$10$^{15}$ cm$^{-2}$ and two dots with a dose of 1$\cdot$10$^{12}$
cm$^{-2}$ with the same conditions as for S3.

\section{Optical Characterisation }

For a quantitative characterisation of the samples and to check the
success of the implantation, we combine confocal scanning microscopy
with optical spectroscopy. To allow scanning of complete samples,
with length scales of a few mm with high resolution, we combined a
piezoelectrical nanopositioning system (NP) with a motorized micropositioning
stage (MP).

\subsection{Setup}

\begin{figure}[htbp]
\includegraphics[width=1\columnwidth]{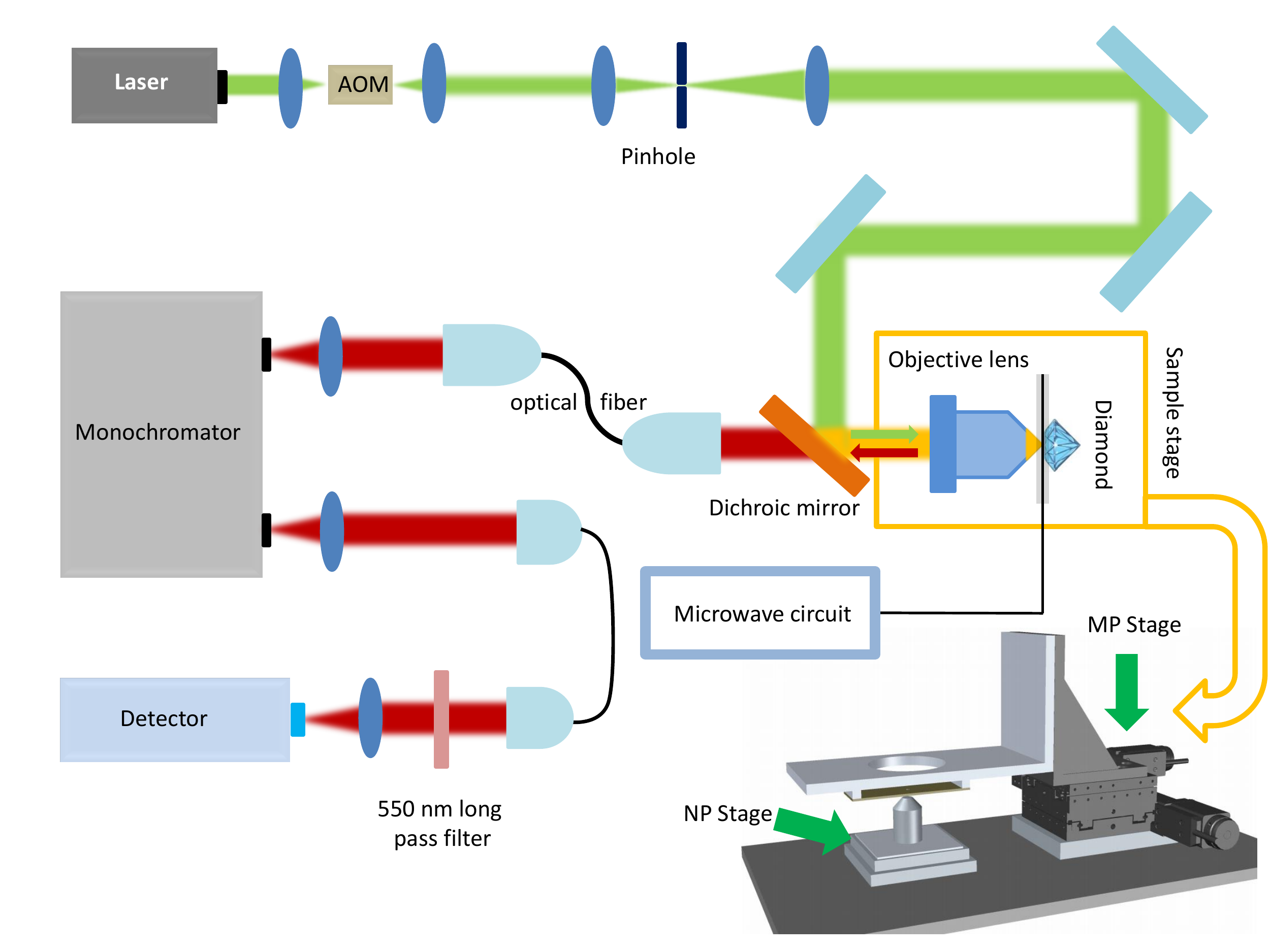}\caption{Schematic diagram of the home-built confocal microscope. It includes
a monochromator and a microwave circuit for exciting the electron
spin transitions of the NV centers. The bottom right part of the figure
shows a 3D representation of the positioning system, which combines
a nanopositioning (NP) stage with a micropositioning (MP) stage to
allow large area scans with high resolution.}
\label{fig:A-schematic-diagram} 
\end{figure}

Fig. \ref{fig:A-schematic-diagram} shows a schematic representation
of the corresponding setup. A diode-pumped solid-state CW laser, which
emits at 532 nm, is used for excitation. An acousto-optical modulator
(extinction ratio 57 dB) is used to switch the laser beam on and off.
The laser beam subsequently passes through an oil immersion microscope
objective (MO) of numerical aperture 1.4, which focuses it into the
diamond sample. The PL signal, propagating in the opposite direction
of the laser beam, is collected by the same MO and separated from
the scattered laser light by the dichroic mirror. The transmitted
PL signal is collected into an optical fiber, passes through a 550
nm long pass filter and is measured by a photon-counting detector.
Alternatively, it can be sent through a monochromator for spectral
analysis. The spot size diameter of our confocal set-up was 0.46 \textgreek{m}m.
The optical fibers in the detection part provide the flexibility to
switch between different types of measurements. The signal from a
MW signal generator (APSIN) and an arbitrary waveform generator (AWG)
are combined to generate the MW signal for exciting the electron spin
transitions of the NV centers. A switch generates MW pulses, which
are passed through a 16 W amplifier and a Cu wire attached to the
diamond sample.

The confocal microscope combines a nanopositioning (NP) stage with
a maximum traveling range of 70 $\mu$m $\times$ 70 $\mu$m in the
XY plane and 50 $\mu$m along the Z-direction with a motorized micropositioning
(MP) stage, as shown in the bottom right of Fig. \ref{fig:A-schematic-diagram}.
The MO is attached to the NP stage whereas the sample is attached
to the MP stage. With this setup, we can scan areas of up to 25 $\times$
25 mm and generate PL images of XY planes at different depths covering
the full diamond samples with nm resolution.

\subsection{Confocal Scanning Microscopy}

\begin{figure}
\includegraphics[width=1\columnwidth]{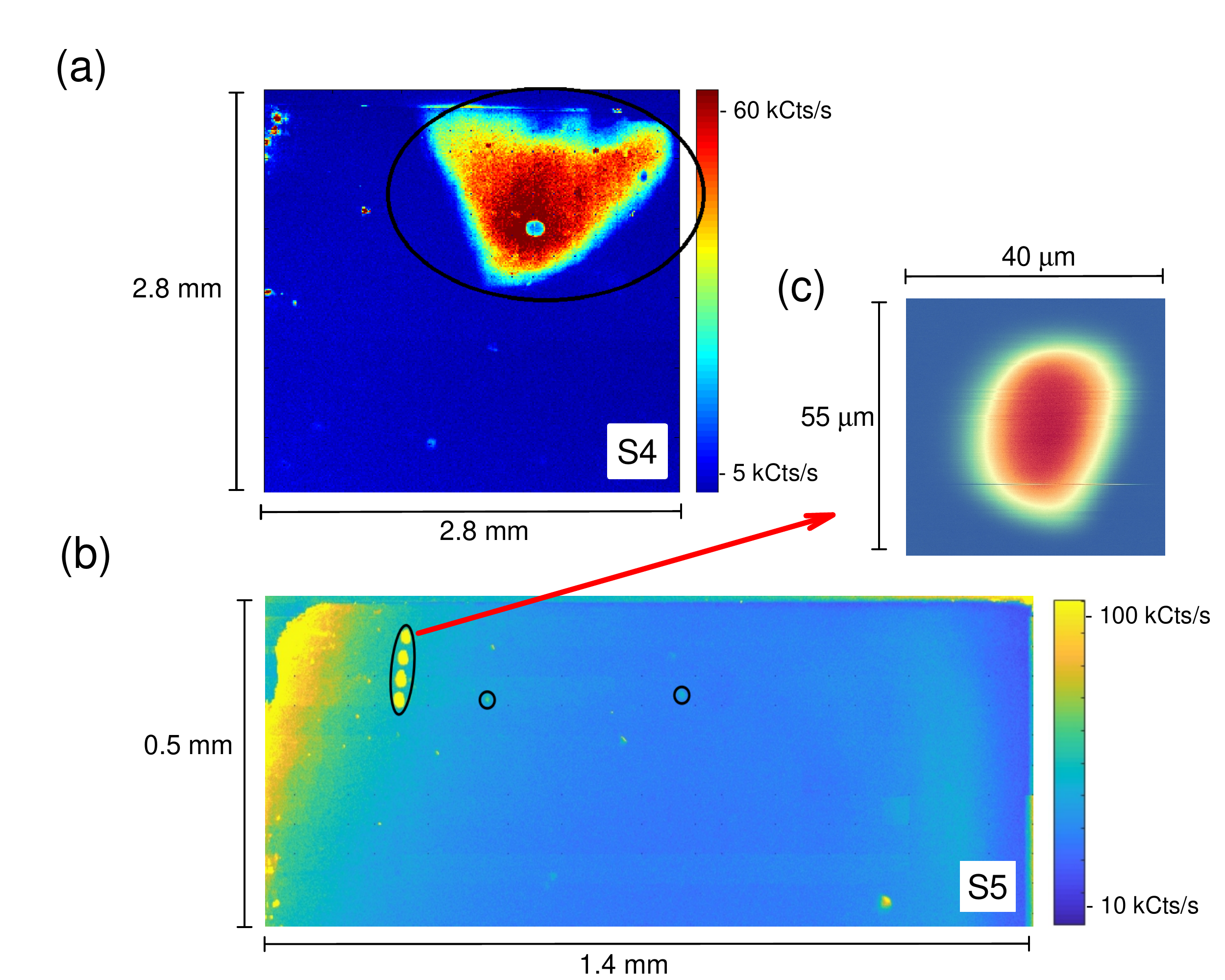}\caption{(a) and (b) show the full area confocal scan images for samples S4
and S5 (details mentioned in section III, specifically in table \ref{Implantation_conditions}).
The ovals mark the regions where ions were implanted to generate NV
centers. The color bars on the RHS of the images indicate the corresponding
count rates. (c) shows an enlarged view of one of the implanted spots
as shown by the arrow. }
\label{area_scan_spectrum} 
\end{figure}

Fig. \ref{area_scan_spectrum} (a) and (b) shows the measured full
scans of the surfaces for samples S4 and S5 respectively. The important
parameters used for preparing S4 and S5 can be found in table \ref{Implantation_conditions}.
The 3 regions marked by ovals in S5 were irradiated with the focused
ion beam to generate NV centers; they clearly show the largest fluorescence
rates. The parameters for implantation are given in section III. Fig.\,\ref{area_scan_spectrum}(c)
shows that the spot size ($\Delta x=15\mu$m, $\Delta y=27\mu$m)
is quite similar to the aperture size of the ion gun. In order to
standardize the preparation conditions, instead of preparing the ensembles
of centers throughout a diamond substrate, with different implantation
parameters like temperature, heating time, ion dose etc., we implanted
several NV spots on a single sample and characterized them. This saved
both the time and the cost of the diamond substrates. Through these
implantations, we also gained better control on precisely creating
spots with NVs. Such control is necessary for many quantum technological
applications. For example, in case of electrical spin read-out, the
NVs should be precisely created in certain positions between the electrodes
\citep{siyushev2019photoelectrical}. Fig. \ref{area_scan_spectrum}(c)
shows an enlarged view of one of the four implanted spots marked by
the oval as shown by the arrow. From the mapping image shown in Fig.
\ref{area_scan_spectrum}(c), we can conclude that the size and position
of the implanted area can be controlled in a deterministic way. The
PL mapping also indicates that the deposited diamond layer appears
to be quite clean. The spectroscopic characterization of this fluorescent
spot is described in section \ref{subsec:OptSpectroscopy}.

The blue area of the sample S4 corresponds to count rates of $\approx$
5000 s$^{-1}$, which is close to the background rate of our system.
In this region, the diamond is very clean, with a very low defect
density.

\begin{figure}[h]
\includegraphics[width=1\columnwidth]{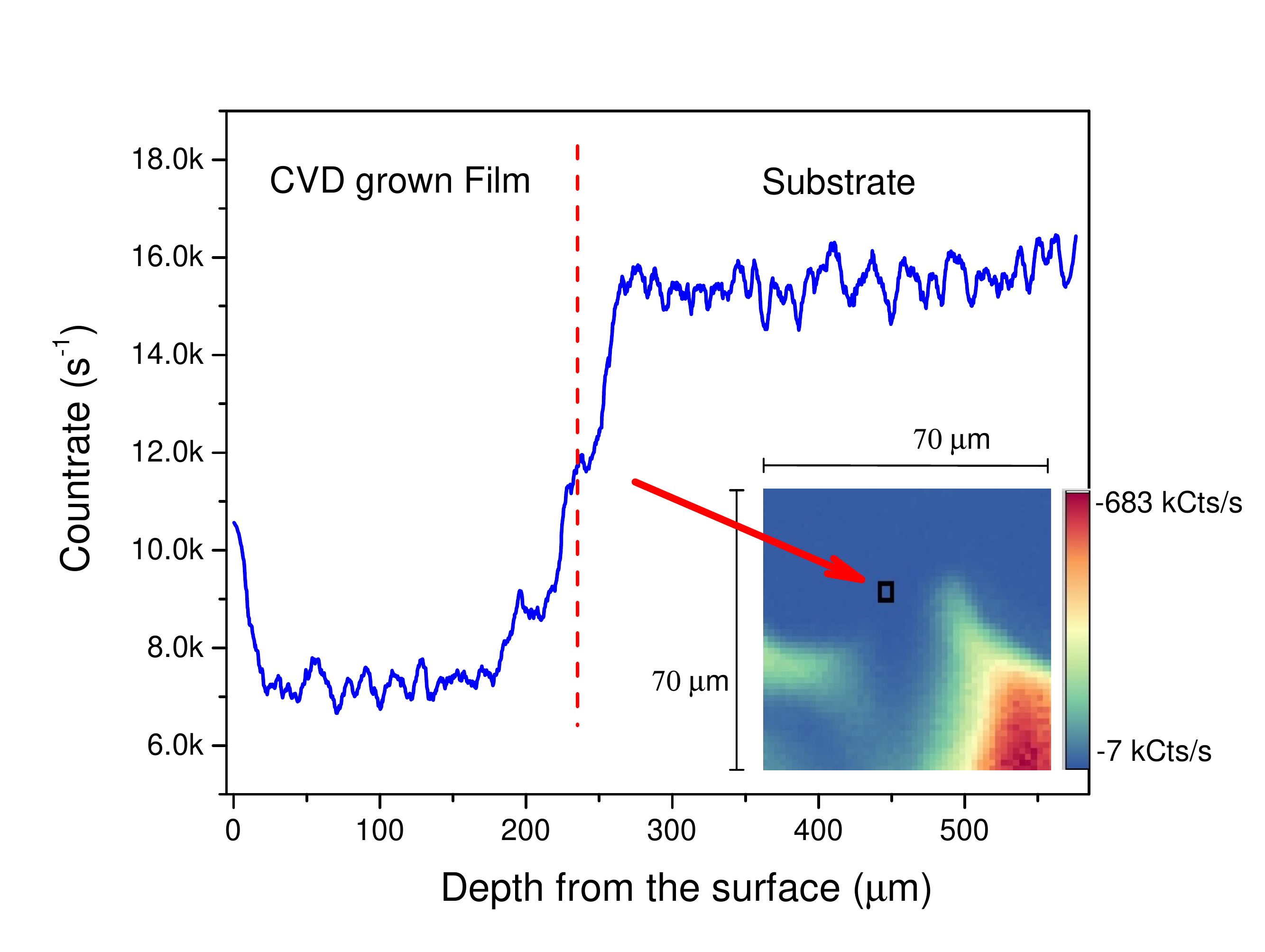}\caption{Depth profile for CVD-grown diamond film on an electronic grade substrate
from Element Six. The vertical line marks the boundary between the
CVD-grown film and the substrate. The inset indicates the scanned
image of the position where the depth scan was performed. \label{fig:The-depth-profile}}
\end{figure}

We used similar measurements to characterise the sample along the
direction perpendicular to the surface. Since the thickness of the
grown films is a few hundred micrometers, we again combined the MP
and NP stages to provide sufficient traveling range. We performed
spatially resolved PL measurements where the focal position of the
MO was varied along the Z-direction through the film. Fig. \ref{fig:The-depth-profile}
shows a depth-profile measured at the position indicated in the inset.
The data indicates that the film thickness is around 265 $\mu$m.
The recorded count-rate in the grown layer is close to the background
rate and is less than the count-rate in the substrate, which indicates
that the CVD-grown film has significantly higher purity than the substrate.
The relatively thick diamond films of 265 \textmu m were grown to
have high purity \textquotedblleft bulk-like\textquotedblright{} diamond
films for the implantation of nitrogen. We wanted to ensure that the
impurities in the diamond substrate do not affect the investigations
of the NV centers on the grown layer. However, the deposition process
is very customizable and films can be grown as thin as a few nm as
well as up to mm thickness. On the other hand, the distance of the
NV centers from the surface is determined by the energy of the nitrogen
ions. The energy range of the ion gun goes from 400 eV up to 5 keV
resulting in a depth of the NV centers between sub nm scale and 10
nm below the surface of the grown layer. For our samples the depth
was between 7 and 10 nm. 

\subsection{Optical Spectroscopy}

\label{subsec:OptSpectroscopy}

\begin{figure}[b]
\centering \includegraphics[width=1\columnwidth]{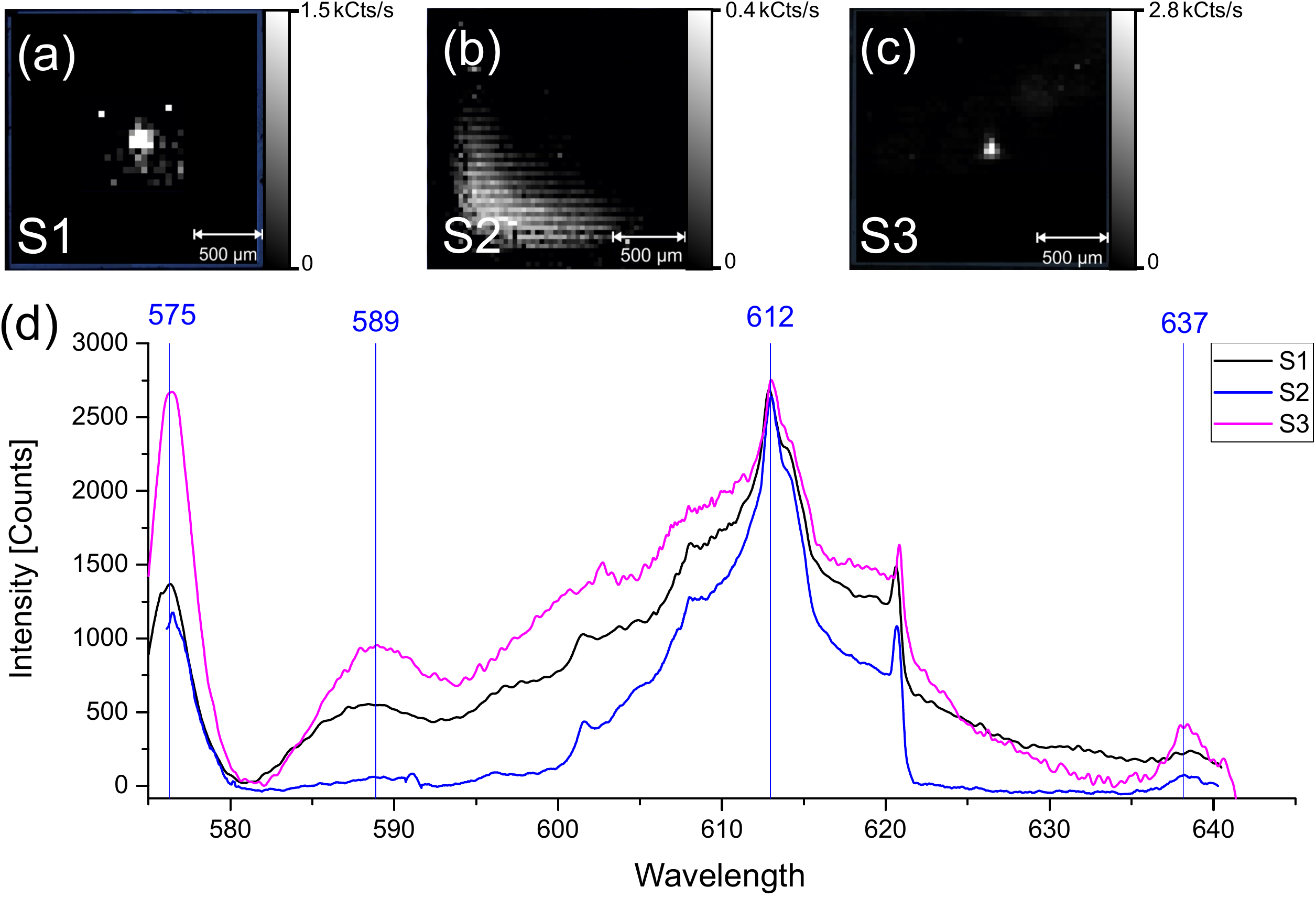}
\caption{2x2\,mm$^{2}$ overview maps of the diamond samples S1 (a), S2 (b)
and S3 (c). The contrast is given by the counts of the NV$^{0}$ ZPL
at 575\,nm (a and c) or NV$^{-}$ ZPL at 637\,nm (b). The spectra
at the bottom are taken from the pixels with the highest intensity.}
\label{fig:spectra1} 
\end{figure}

\begin{figure}[h]
\centering \includegraphics[width=1\columnwidth]{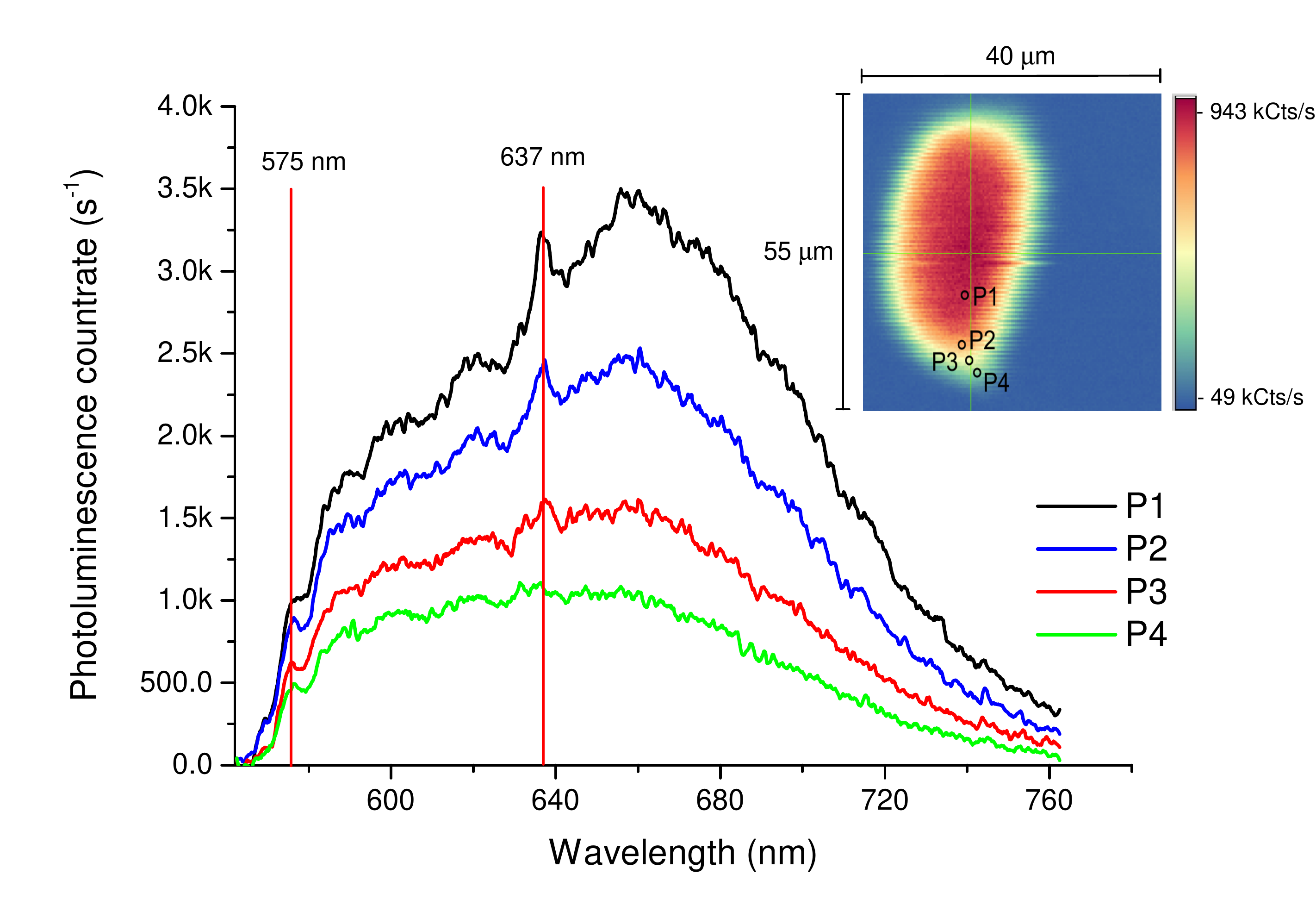} \caption{4 spectra taken at different positions near the implantation spot
for sample S5. The positions range from an area of strong signal near
the center to the edge, where the count rate is low. The inset shows
the spatial distribution of the PL intensity. }
\label{fig:spectra2} 
\end{figure}

The successful formation of NV centers in the grown diamond films
was confirmed by measuring spectra of the the fluorescence emitted
by these regions. For each experiment a 2$\times$2\,mm$^{2}$ map
with 50$\times$50\,pixel$^{2}$ was done with a \textit{Renishaw
Raman Spectrometer} as a simple spatially resolved spectrometer. At
each pixel we recorded a PL-spectrum. Fig. \ref{fig:spectra1} (d)
shows some of these spectra. The peaks in the spectra are associated
with NV$^{0}$ ZPL at 575\,nm (curves 1 and 3) or NV$^{-}$ ZPL at
637\,nm (2). An overview of the position and shape of the implanted
areas was obtained using the mapping method described above. The details
about the samples are mentioned in table \ref{Implantation_conditions}.
The map of sample S1 in Fig. \ref{fig:spectra1}(a) shows a spot at
the center of the diamond with a diameter of $\sim$200\,$\mu$m
and a halo of $\sim$400\,$\mu$m, which is the spot size that we
expected from our experimental condition for implantation: the ion
gun was operated with an emission current $\sim$ 100 $\mu A$. Using
smaller apertures, we obtained smaller spots, such as the one shown
in Fig. \ref{area_scan_spectrum}(c) , which has a diameter of $\sim$25\,$\mu$m
, obtained with an ion beam current of 500 pA. The spectrum for S1
shows the zero phonon lines (ZPL) of NV$^{0}$ at 575\,nm and of
NV$^{-}$ at 637\,nm, the second order Raman peak between 600 - 620
nm, and a peak at 589\,nm that is associated with some point defects
possibly created during ion implantation\citep{zaitsev2013optical}.
In Fig. \ref{fig:spectra1} (b) S2 shows a strong enlargement of the
implantation region on the diamond surface to the lower left corner
of the sample. As described in section III, this implantation was
performed with a non-terminated and therefore electrically insulating
diamond surface. Accordingly, the charges deposited by the beam result
in surface charging and subsequent defocussing of the beam. This shows
that the bare diamond surface at room temperature is not suited for
localized nitrogen implantation at the low energies required for shallow
NV centers. The peak for NV$^{-}$ has a lower count rate than in
the sample S1, which indicates a lower density in the laser stimulated
area, caused by the broadening of the beam. The spectrum for S3 shown
in Fig. \ref{fig:spectra1} (c) is similar to that of S1 but has slightly
stronger features. S3 shows a sharp spot of $\sim$150\,$\mu$m diameter
without a bright halo. Using the relative ZPL intensities of NV$^{0}$
and NV$^{-}$ charge states for S1, S2 and S3, we estimated their
approximate concentration ratios $C_{NV^{0}}:C_{NV^{-}}$of the measured
spots following the method described in ref. \citep{Acousta_PhysRevB.80.115202}.
We obtained that $C_{NV^{0}}:C_{NV^{-}}$ equals 0.71, 2.8 and 1.5
for S1, S2 and S3 respectively.

Fig. \ref{fig:spectra2} exhibits a set of 4 spectra measured using
the monochromator integrated with the confocal set-up (as described
in section IV(A)) at different positions ranging form close to the
center of the implanted area towards the periphery, as shown in the
inset. The data clearly show the zero phonon lines (ZPL) at 575 nm
and 637 nm associated with the NV$^{0}$ and NV$^{-}$ centers respectively,
and the corresponding phonon side bands.

\section{Spin Properties}

In this section we describe measurements of the coherence properties
of the NV centers by electron spin-dependent fluorescence measurements
at room temperature. The negatively charged NV center forms a triplet
ground state where the zero field splitting separates the $m_{S}=0$
and $\pm1$ states by $\sim$2.87 GHz \citep{Doherty:2013uq,suter2017single}.
An external magnetic field lifts the degeneracy of the $m_{S}=\pm1$
levels and of the ODMR lines. As an ensemble of NV centers consist
of four possible symmetry axes, an applied magnetic field along an
arbitrary direction projects itself onto the four axes with four different
field values which creates four pairs of electronic transitions and
hence eight lines in the ODMR data \citep{pham2013magnetic}. In our
experiments, the orientation of the applied field was along such a
direction that the components of the applied magnetic field along
the four possible NV axes of the ensemble we measured had similar
values, resulting in only 2 lines in the ODMR spectrum. The magnitude
of the applied magnetic field was 16 Gauss. By scaling the photoluminescence
with respect to the signal from a single center we estimated the concentration
\ensuremath{\aleph} of the of the NV centersin the spot which we performed
the ODMR measurements as \ensuremath{\aleph}=22 ppm. To measure the
coherence properties, we used different experiments that first initialized
the system into the bright state ($m_{S}=0$) by a 5$\mu s$ laser
pulse. From this state, the coherence is generated and manipulated
by resonant microwave pulses and the final state is read out by counting
fluorescence photons during a second laser pulse whose optimal duration
is $\sim400$ns \citep{suter2017single}.

\begin{figure}
\includegraphics[width=1\columnwidth]{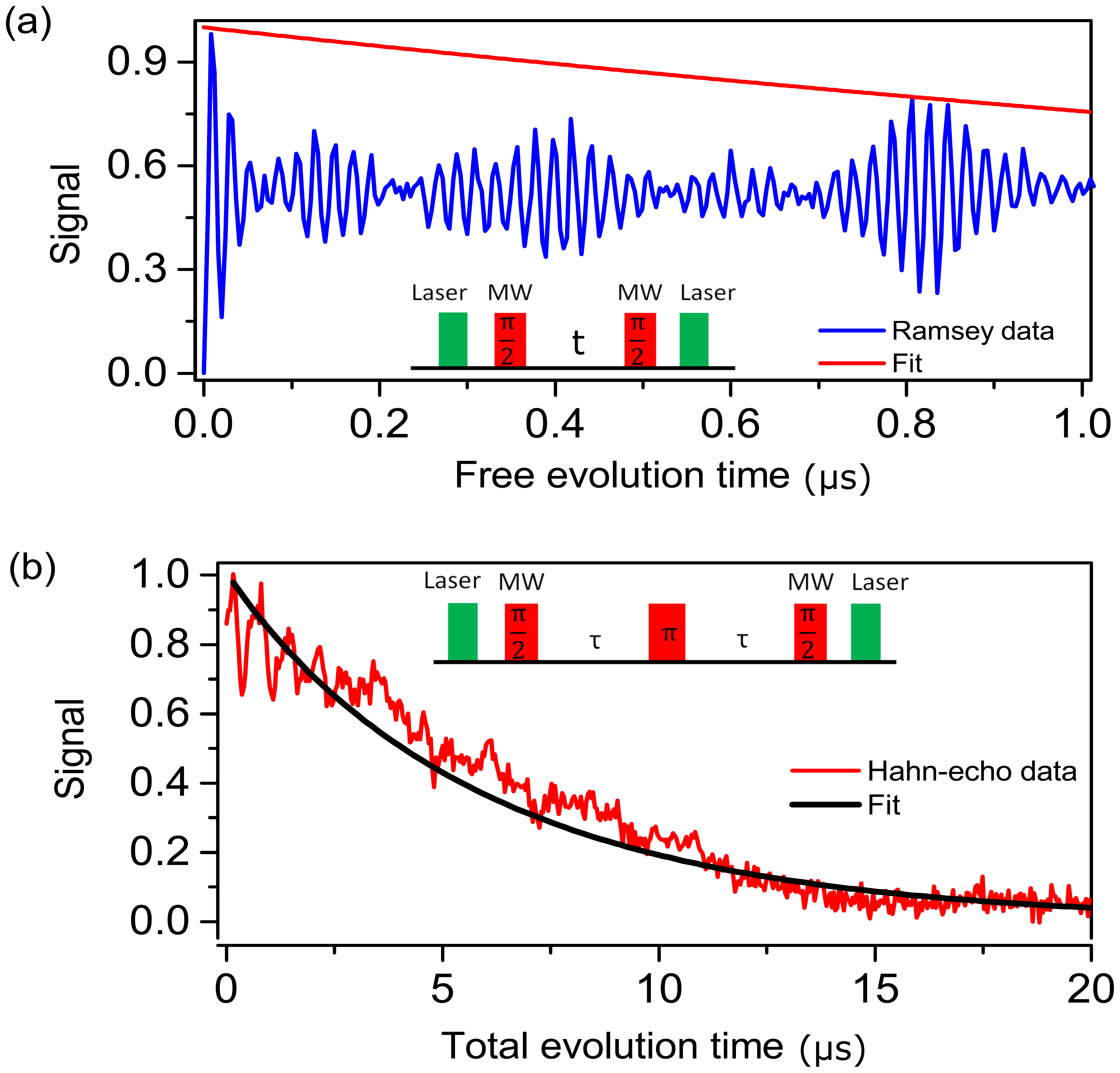}\caption{(a) Free induction decay of the NV spins and (b) decay of the coherence
of an ensemble of NV spins measured using the Hahn-echo sequence.
The corresponding pulse sequences are shown as insets. }
\label{FID_HahnEcho} 
\end{figure}

Fig. \ref{FID_HahnEcho}(a) shows the resulting free-induction decay
(FID) signal, measured with a Ramsey-type \citep{ramsey1950molecular}
experiment where two $\frac{\pi}{2}$ pulses separated by a free evolution
period $t$ were applied between the initialization and read-out laser
pulses. The first $\frac{\pi}{2}$ pulse, with a frequency of 2.8272
GHz, generates a superposition of the spin states $m_{S}=0$ and $-1$,
which subsequently evolves for a time $t$. The second $\frac{\pi}{2}$
pulse converts one ocmponent of the coherence back into population
which is measured during the read-out step. The observed oscillation
frequency of 50 MHz is the difference between the carrier frequency
of the microwave between the two pulses and the transition frequency.
The beats are due to the hyperfine coupling between the electron spin
and the $^{14}$N nuclear spin. We fitted the envelope of the FID
signal to the expression $e^{-t/T_{2}^{*}}$ , and calculated $T_{2}^{*}$=3.6$\mu$s.

To measure the coherence time $T_{2}$, we used the spin-echo sequence
introduced by E. L. Hahn \citep{hahn1950spin}. The sequence starts
with the 5$\mu s$ initialization laser pulse followed by three MW
pulses separated by free precession periods $\tau$: $\frac{\pi}{2}-\tau-\pi-\tau-\frac{\pi}{2}$
. The first $\frac{\pi}{2}$ pulse creates a coherent superposition
of the $m_{S}=0$ and $-1$ states and the last one converts the coherence
into population difference. The $\pi$ pulse inverts the accumulated
phase, resulting in zero overall phase after the second free precession
period, provided the environment is static. The experimental data
in Fig. \ref{FID_HahnEcho}(b) show the decay of the coherence of
the ensemble of NV spins. The envelope of the decay curve is fitted
to the expression $e^{-(t/T_{2})^{p}}$, where $t=2\tau$ and the
exponent $p$ depends on the magnetic environment \citep{bar2012suppression,pham2012enhanced}.
We obtained $T_{2}$(Hahn echo)=$6.4\mu s$ and $p=0.96$. 

Preserving the coherence of quantum states on a longer timescale is
essential for quantum information and sensing protocols \citep{ladd2010quantum,Maurer1283,putz2014protecting}.
Decoherence of the spins belonging to NV centers happens due to their
undesired interaction with the magnetic environment mainly caused
by the electronic spins of substitutional nitrogen centers and $^{13}\mathrm{C}$
nuclear spins \citep{de2010universal,de2011single,naydenov2011dynamical,ryan2010robust}.
To protect quantum states from decoherence, dynamical decoupling (DD)
has been established as an efficient technique \citep{souza2012robust,RevModPhys.88.041001}.
It decouples the system spins from the surrounding spin-bath by applying
a periodic sequence of inversion pulses that effectively isolate the
system spin from the environmental noise \citep{de2010universal,bar2013solid,pham2012enhanced,naydenov2011dynamical,bar2012suppression}.

\begin{figure}
\includegraphics[width=1\columnwidth]{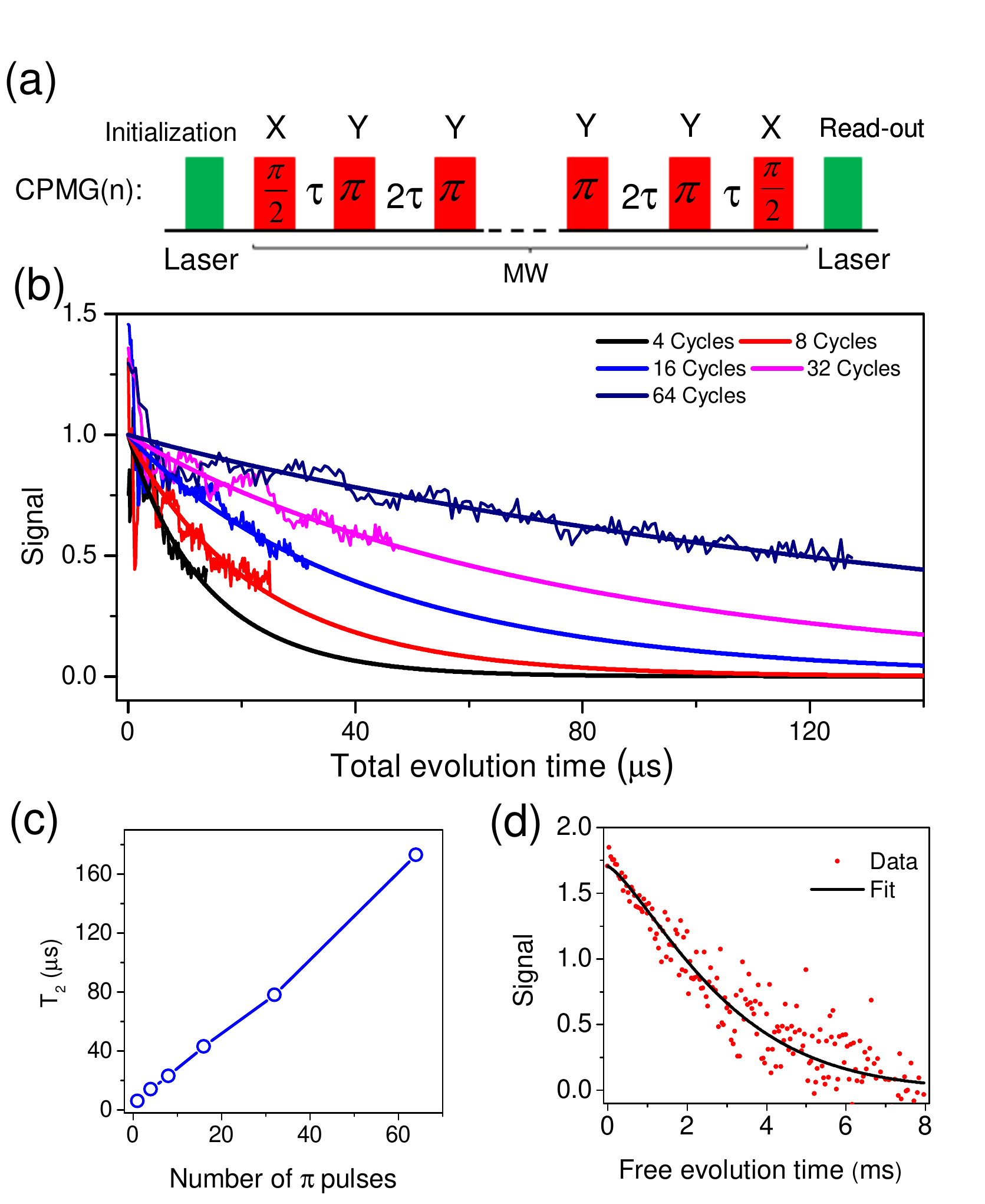}\caption{(a) CPMG control sequence for variable number of refocusing $\pi$
pulses. (b) Measured coherence decay data for an ensemble of NV spins
for 4, 8, 16, 32 and 64 number of $\pi$ pulses. The signal is in
arbitrary units. The thick solid lines represent the fitted curves.
(c) Variation of $T_{2}$ with the number $n$ of CPMG $\pi$ pulses.
(d) $T_{1}$ relaxation for the electronic spin.}
\label{CPMG} 
\end{figure}

Here, we demonstrate that the electronic spins in our samples respond
well to the dynamical decoupling sequences and we are able to extend
the coherence time $T_{2}$ beyond the values obtained in the Hahn-echo
measurements described above. For this purpose, we use the Carr\textendash Purcell\textendash Meiboom\textendash Gill
(CPMG) pulse sequence, a robust DD technique which has been extensively
used in NMR spectroscopy \citep{souza2012robust}. The CPMG sequence
can be considered as an extended version of the Hahn-echo sequence
where multiple refocusing $\pi$ pulses are applied during the free
evolution period between the two $\frac{\pi}{2}$ pulses, as shown
in Fig. \ref{CPMG}(a). We implemented CPMG sequences with the number
$n$ of control pulses varying from 4 to 64. The measured decoherence
curves as a function of total free evolution time $t=2n\tau$ and
their fit to the expression $e^{-(t/T_{2})^{p}}$ are plotted in Fig.
\ref{CPMG}(b). We have determined the coherence time $T_{2}(n)$
for the ensemble of NV spins from the above analysis and observed
its extension by almost 29 times for 64 pulses {[}$T_{2}(64)=173\mu s${]}
over the corresponding value measured by the Hahn-echo. The variation
of $T_{2}$ with $n$ is shown in Fig. \ref{CPMG}(c). We also measured
the spin-lattice relaxation time for the same ensemble of NV centers.
Fig. \ref{CPMG}(d) shows the measured data along with the theoretical
fit to the expression $e^{-(t/T_{1})^{q}}$. We obtained $T_{1}=3.14$
ms and $q$=1.32. For short echo times, a part of the signal decays
more rapidly. This appears to be due to some initial transients that
precess around the effective field but die out after a few microseconds.

\begin{figure}
\includegraphics[width=1\columnwidth]{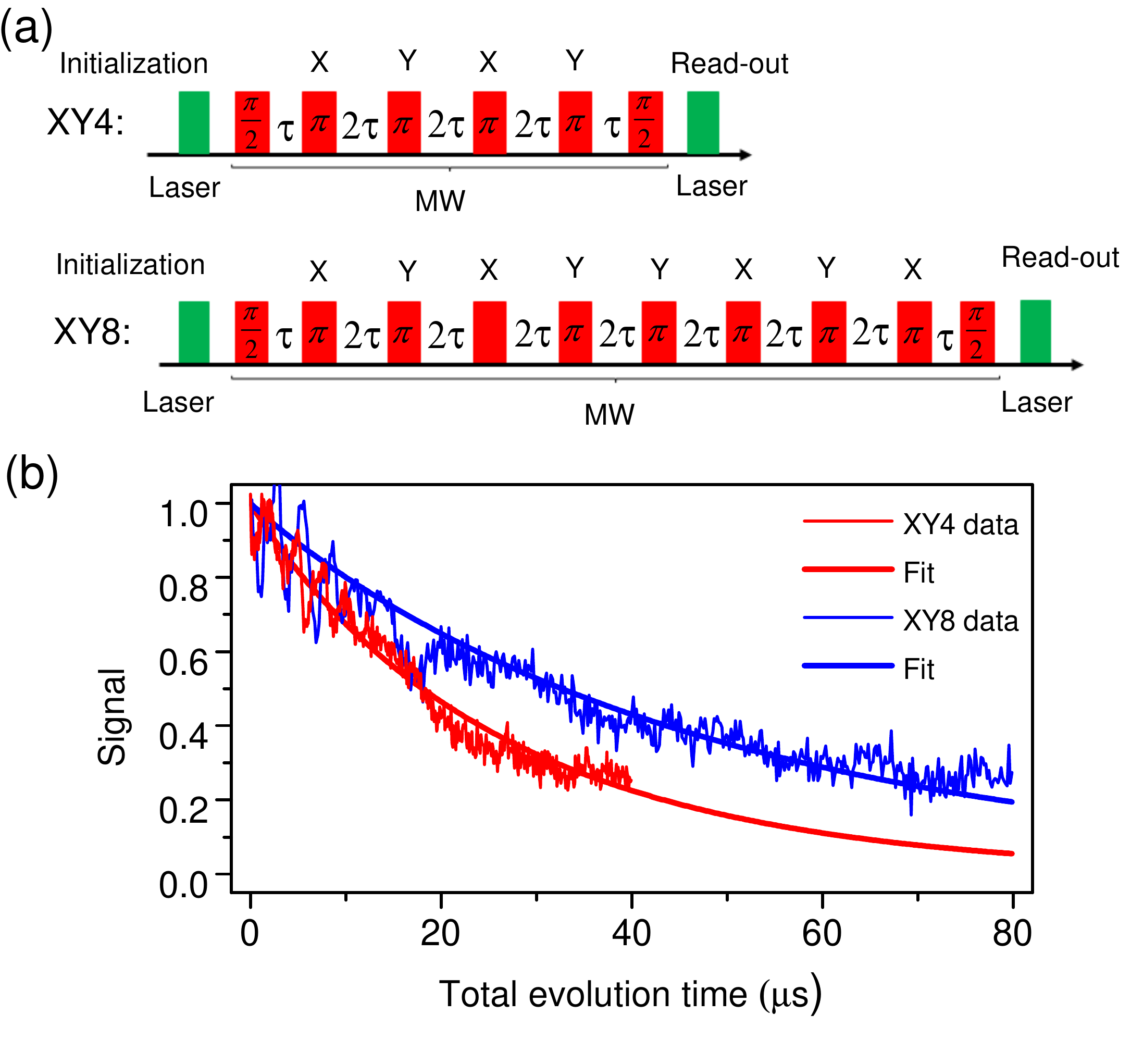}\caption{(a) XY4 and XY8 dynamical decoupling sequences. (b) Coherence decay
curves for multiple NV spins measured using XY4 and XY8 control sequences
along with their theoretical fits to the expression $e^{-(t/T_{2})^{p}}$where
$t$ equals $8\tau$ and $16\tau$ for the XY4 and XY8 sequences respectively. }
\label{XY4_XY8} 
\end{figure}

While the CPMG sequence is well suited for preserving a given quantum
state, it is rather unsuited for protecting unknown quantum states
\citep{Maudsley1986488,PhysRevLett.106.240501}. For such cases, it
is therefore important to use a pulse sequence that performs well
for arbitrary initial conditions. We therefore repeated the measurement
with the pulse sequences XY4 and XY8. They are well established as
robust and symmetrized sequences \citep{Maudsley1986488,PhysRevLett.106.240501,de2011single,ryan2010robust}.
They consist of a series of $\pi$ pulses around two perpendicular
rotation axes as shown in Fig. \ref{XY4_XY8}(a). The experimental
decay curves obtained with the XY4 and XY8 sequences are shown in
Fig. \ref{XY4_XY8}(b) together with the corresponding fits. We were
able to extend $T_{2}$ up to $26.4\mu s$ and $47.8\mu s$ by using
the XY4 and XY8 sequences, respectively.

The sensitivity of a spin based magnetometer is limited by the photon
shot-noise when the readout is performed optically. An effective way
to enhance the shot-noise limited sensitivity is to use a large number
of NV spins which can enhance the collected PL signal significantly.
Moreover, reduced dephasing and an enhancement in the coherence time
of the NV spins can directly improve the sensitivity. Here, we try
to overcome these issues by creating ensembles of NV centers with
superior coherence properties and examine the sensitivity of our possible
NV based magnetometer. Thus, following the speculation of ref. \citep{taylor2008high},
we can conclude that in ideal measurement conditions for an NV-based
magnetometer exploiting the NV centers we prepared, an NV concentration
$\aleph=22$ ppm and coherence time $T_{2}^{*}=3.6\mu s$ can give
a DC magnetic field sensitivity $\eta_{dc}$ up to $\sim$100 nT Hz$^{-1/2}$
. On the other hand, by appying the DD sequences we could enhance
the coherence time $T_{2}$(DD) significantly which can increase the
sensitivity $\eta_{ac}$ to ac magnetic field by a factor of $\sqrt{T_{2}(DD)/T_{2}^{*}}.$
For instance, using the DD sequence CPMG(64) we can obtain $\eta_{ac}\sim$10
nT Hz$^{-1/2}$ .

\section{Conclusion}

Through this project we have described the deterministic preparation
of ensembles of NV centers in ultra-pure diamond films through three
major steps: growing the diamond films, implanting the NV centers
and characterizing them. A plasma assisted CVD reactor for depositing
high purity diamond films was constructed and optimized to minimize
the nitrogen content in the synthesized films. High-purity films with
a thickness of 260 \textgreek{m}m at a growth rate of about 30 \textgreek{m}m/h
with constant high crystal quality were deposited. The nitrogen content
of the deposited diamond layers is below the detection limits of the
standard methods of analytical chemistry or physics. The upper limit
for nitrogen in our films was measured to be 1.3$\cdot$10$^{-7}$(ca.
0.1 ppm) which refers that our films are at least among the purest
diamonds substrates that are commercially available. In the future
we plan to perform measurements like EPR to see whether these samples
are having even better purity than the commercial available ones.

A novel experimental set-up for in situ preparation of the diamond
surface, implantation and high temperature annealing has been presented.
In conclusion it was observed that it is possible with this new method
to create NV centers while heating the diamond in UHV without strong
defocussing effects. It was shown that it is possible to create localized
NV centers on hydrogen terminated and heated diamonds. Optimized implantation
parameters have been obtained which enabled us to precisely implant
NV centers. Using the above described implantation through an aperture,
we could implant spots with sizes similar to the aperture of the ion
gun in a deterministic way, thus achieving better a better localization
of the NV centers. The experimental protocol of large area scanning
has been demonstrated which enables us to record the PL image of the
full sample in mm scale with nm resolution. Optical spectroscopic
measurements have ascertained the generation of NV centers. Next,
we have demonstrated the potentiality for possible application of
the created NV centers in quantum information and sensing by preserving
the coherence for extended times. We observed that the NV spins respond
well to different robust dynamical decoupling sequences. The coherence
time $T_{2}$ has been extended and dephasing of he spins has been
delayed in a controlled fashion. We can conclude that a magnetic sensor
designed using the sample we prepared can reach a sensitivity of $\sim$10
nT Hz$^{-1/2}$ under optimized experimental conditions. 

In the future we plan to improve the coherence properties of the spins
by creating NV centers in isotopically purified $^{12}$C enriched
diamond with $^{13}$C concentration below $0.01\%$ and improve the
sensitivity to ac and dc magnetic field. Moreover, we plan to design
an ion gun with nm aperture and install a single ion counter which
will allow us to optimize the implantation dose and hence, the NV
concentration. These experimental upgrades will allow us to deterministically
implant ordered arrays of single NV centers which can function as
an efficient large-scale solid-state quantum register.
\begin{acknowledgments}
We acknowledge K. R. K. Rao for his assistance with developing the
software for area scanning experiments. The research under this project
was supported by the funding from Mercator Research Center Ruhr (MERCUR). 
\end{acknowledgments}

\bibliographystyle{apsrev4-1}
\bibliography{draft}

\end{document}